\documentclass[12pt]{article}

\usepackage{graphicx}
\usepackage{amsmath}
\usepackage{amssymb}
\usepackage{amsthm}
\usepackage{pstricks}
\usepackage{color}
\usepackage{here}

\renewcommand{\thefootnote}{\fnsymbol{footnote}}

\begin{document}

\begin{flushright}
EPHOU-16-002 \\
MISC-2016-02
\end{flushright}

\vspace{4ex}

\begin{center}

{\LARGE\bf Dilaton Stabilization in Three-generation Heterotic String Model}

\vskip 1.4cm

{\large  
Florian Beye$^{1}$\footnote{Electronic address: fbeye@eken.phys.nagoya-u.ac.jp},
Tatsuo Kobayashi$^{2}$\footnote{Electronic address: kobayashi@particle.sci.hokudai.ac.jp}
and
Shogo Kuwakino$^{3}$\footnote{Electronic address: kuwakino@cc.kyoto-su.ac.jp}
}
\\
\vskip 1.0cm
{\it $^1$Kawasaki, Japan} \\
{\it $^2$Department of Physics, Hokkaido University, Sapporo 060-0810, Japan} \\
{\it $^3$Maskawa Institute for Science and Culture, Kyoto Sangyo University, Kamigamo-Motoyama, Kita-ku, Kyoto 603-8555, Japan} \\

\vskip 3pt
\vskip 1.5cm

\begin{abstract}
We study dilaton stabilization in heterotic string models. By utilizing the asymmetric orbifold construction, we construct an explicit three-generation model whose matter content in the visible sector is  the supersymmetric standard model with additional vectorlike matter. 
This model does not contain any geometric moduli fields except the dilaton field. Model building at a symmetry enhancement point in moduli space enlarges the rank of the hidden gauge group. By analyzing multiple hidden gauge sectors, the dilaton field is stabilized by the racetrack mechanism. We also discuss a supersymmetry breaking scenario and F-term uplifting. 
\end{abstract}

\end{center}

\newpage

\setcounter{footnote}{0}
\renewcommand{\thefootnote}{\arabic{footnote}}

\section{Introduction}

Superstring theory provides a comprehensive way to unify a model of particle physics and gravitational theory in a quantum theory. Three gauge interactions of the standard model of particle physics and the graviton, which arises from quantization of gravity, are realized as massless string excitation modes. The six-dimensional extra dimensions which are predicted by superstring theory are compactified into some manifold, and strings  on the manifold provide geometrical interpretations of fundamental particles --- quarks and leptons.

In order to make string-derived models phenomenologically realistic, moduli stabilization and supersymmetry (SUSY) breaking in the low energy effective theory should be addressed. 
These issues have been studied extensively in D-brane string models (see for a review \cite{Ibanez:2012zz}).
For example, in the Kachru-Kallosh-Linde-Trivedi scenario of type IIB orientifolds \cite{Kachru:2003aw}, complex structure moduli fields as well as the dilaton field 
are stabilized by inclusions of 3-form fluxes among compactified internal spaces, and K\"ahler moduli fields can be 
stabilized by non-perturbative effects caused by hidden sector dynamics.
At that stage, there is a supersymmetric anti de Sitter vacuum. Then, 
 SUSY is explicitly broken by inclusion of anti-D-branes, and such SUSY breaking uplifts the vacuum to lead to 
very small but positive vacuum energy.
Spontaneous SUSY breaking can also uplift the vacuum, e.g. in the F-term uplifting scenario 
\cite{Lebedev:2006qq,Dudas:2006gr}.
 However, in heterotic string models, these issues are not solved easily though several attempts for moduli stabilization and SUSY breaking in heterotic string models have been made 
(see for a review \cite{Ibanez:2012zz}).

In heterotic string models, in order to address these issues, non-perturbative effects such as gaugino condensation should be taken into account. Moduli fields can be stabilized by a potential which is made by multiple gaugino condensations --- this is called the racetrack mechanism \cite{Krasnikov:1987jj}. One of the possibilities to break SUSY is dynamical SUSY breaking. In \cite{Intriligator:2006dd}, it is argued that, in $\mathcal{N}=1$ SQCD with a suitable number of light flavors, SUSY is broken dynamically at a non-supersymmetric and long-lived metastable vacuum. 
Moreover, heterotic string theory in generic compactification includes many moduli fields such as the
dilaton, K\"ahler moduli and complex structure moduli fields.
We need several dynamics to stabilize all of them.
However, realization of  suitable hidden gauge sectors which cause gaugino condensations is strongly restricted in heterotic string models since the rank of gauge symmetry is typically 16. Our standard model gauge group $SU(3)_{{\rm C}} \times SU(2)_{{\rm L}} \times U(1)_{{\rm Y}}$ has rank 4, and heterotic string models have multiple additional $U(1)$s (commonly around ten, though this number is model dependent of course). Thus there is a little room for multiple hidden gauge sectors.

Our strategy to attack this problem is  to start with a string model at a symmetry enhancement point in moduli space in order to enlarge a rank of hidden gauge groups. In this paper, we consider moduli stabilization in heterotic string models from the asymmetric orbifold construction \cite{Narain:1986qm}. In this formalism, the rank of gauge symmetry is enhanced: it attains the maximum allowed value of 22, since we make models from a Narain lattice that corresponds to a symmetry enhancement point in moduli space. Thus we expect that there are several hidden gauge sectors which are suitable for moduli stabilization and SUSY breaking. Furthermore, by the asymmetric orbifold action, geometric moduli fields except the dilaton field are frozen \cite{Mueller:1986yr}. Then the number of moduli fields is very small, and we can concentrate on dilaton stabilization only. In this paper, dilaton stabilization and SUSY breaking by the enlarged hidden gauge sectors will be discussed.

This paper is organized as follows. In Section 2, we construct a three-generation model from a heterotic asymmetric orbifold. In Section 3, we analyze dilaton stabilization in the obtained three-generation model. We also comment on a possible SUSY breaking scenario. The last section is devoted to conclusions.

\section{Three-generation asymmetric orbifold model}

In this section, we construct a four-dimensional SUSY standard model with three generations in the framework of the heterotic asymmetric orbifold construction \cite{Narain:1986qm}. In \cite{Beye:2013moa, Beye:2013ola, Ito:2010df}, the model building procedure and some three-generation models are shown. We consider a ${\bf Z}_3$ asymmetric orbifold action, which is the simplest choice for four-dimensional $\mathcal{N}=1$ SUSY. 

Our starting point is a four-dimensional heterotic string theory compactified on a Narain lattice \cite{Narain:1985jj}. Narain lattices can be described by a metric, B-fields and Wilson lines along the internal six-dimensional spaces, which are suitably chosen to be on an enhancement point in the moduli space \cite{Narain:1986am}. Thus the rank of the gauge symmetry is the maximum allowed value of 22, which originates from 22-dimensional compactified directions in the left mover. At the gauge enhancement point we can realize a relatively rich source of multiple hidden sectors.

Here we build a string model from the $A_3^7 \times U(1) \times \overline{E}_6$ lattice, which is described by the following conjugacy class generators
\begin{align}
&(0, 0, 0, 0, 0, 0, 0,1/3, 2),  \nonumber \\
&(0, 1, 1, 2, 0, 0, 1,1/4, 0),  \nonumber \\
&(1, 0, 1, 1, 2, 0, 0,1/4, 0),  \nonumber \\
&(0, 1, 0, 1, 1, 2, 0,1/4, 0),  \nonumber \\
&(0, 0, 1, 0, 1, 1, 2,1/4, 0),
\end{align}
of $A_3^7 \times U(1) \times \overline{E}_6$\footnote{This lattice is labeled as lattice \#30 in \cite{Beye:2013moa}, where several Narain lattices for ${\bf Z}_3$ model building are classified. In \cite{Beye:2013ola}, by using the same lattice, a three-generation model was constructed. In this paper, we choose different choice of ${\bf Z}_3$ shift action.}. Here, the normalization for the $U(1)$ is given by $2 \sqrt{3}$. 
To reduce SUSY from $\mathcal{N} =4$ to $\mathcal{N}=1$, we choose the following ${\bf Z}_3$ twist vector $t_{{\rm R}}$ for the right mover
\begin{align}
t_{{\rm R}} &=
( 0, \frac{1}{3}, \frac{1}{3}, - \frac{2}{3} ).
\end{align}
We do not consider any twist action for the left mover. We choose a shift vector $V_{{\rm L}}$ for the left mover as
\begin{align}
V_{{\rm L}} 
&=
( 0, 
\alpha_1^{A_3} + 2\alpha_2^{A_3},
-\alpha_1^{A_3} - 2\alpha_2^{A_3},
-\alpha_1^{A_3} - 2\alpha_2^{A_3},
\alpha_3^{A_3},
\alpha_3^{A_3},
\alpha_3^{A_3},
0,0
 )/3
,
\label{Shiftvector}
\end{align}
where $\alpha_i^{A_3}$ denotes simple roots of the $A_3$ group. This shift vector satisfies the modular invariance condition for ${\bf Z}_3$ orbifold models,
\begin{align}
3 V_{{\rm L}}^2 \in 2 {\bf Z}.
\end{align}
Hence, this model is a consistent model.

Following the model building procedure in \cite{Beye:2013ola}, we can read off the massless spectrum of the four-dimensional model as given in Table \ref{Tab:Spectrum}. By the ${\bf Z}_3$ shift action \eqref{Shiftvector}, the original gauge symmetry $SU(4)^7 \times U(1)$ associated with the starting $A_3^7 \times U(1) \times \overline{E}_6$ Narain lattice reduces to 
\begin{align}
SU(3)_{{\rm C}} \times SU(2)_{{\rm L}} \times SU(2)^2 \times SU(3)^2 \times SU(4) \times U(1)^{10}.
\label{Gaugesymmetry}
\end{align}
This model contains $12 \times 3$ multiplets in the untwisted sector and $36 \times 3$ multiplets in the ${\bf Z}_3$ twisted sector. The factor $3$ in the twisted sector comes from the degeneracy factor $D=3$. In Table 1, we label all fields as $f_i$ for $i=1 \ldots 48$. $U(1)$ normalizations for $U(1)^{10}$ are taken to be
\begin{align}
U_{1,2,3} &= \frac{2}{\sqrt{3}} Q_{1,2,3}, \qquad
U_{4,6,8} = Q_{4,6,8}, \qquad
U_{5,7,9} = \sqrt{2} Q_{5,7,9}, \qquad
U_{10} = \frac{1}{\sqrt{3}} Q_{10}.
\end{align} 
Among the ten $U(1)$s, we find suitable hyper $U(1)_{\rm Y}$ and anomalous $U(1)_{\rm A}$ symmetries by taking the following linear combinations
\begin{align}
Q_{{\rm Y}}
&= \frac{1}{2\sqrt{3}} U_1 - \frac{1}{2} U_4  - \frac{1}{\sqrt{2}} U_5 + \frac{1}{2} U_6 + \frac{1}{2} U_8 - \frac{1}{\sqrt{2}} U_9, \nonumber \\
Q_{{\rm A}}
&= \frac{1}{5\sqrt{3}} U_1 
- \frac{1}{20\sqrt{3}} U_2 
- \frac{1}{20\sqrt{3}} U_3 
- \frac{1}{20} U_4 
+ \frac{1}{30 \sqrt{2}} U_5
- \frac{1}{10} U_6 
- \frac{1}{15\sqrt{2}} U_7
+ \frac{1}{20} U_8
+ \frac{1}{30\sqrt{2}} U_9 
.
\end{align}
By this choice of the hyper charge, it turns out that this model contains suitable three-generation standard model particles. Additional particles are vectorlike fields with respect to the standard model group $SU(3)_{{\rm C}} \times SU(2)_{{\rm L}} \times U(1)_{{\rm Y}}$ which, at low energy scale, are expected to have effective masses, as well as singlets including candidates of right-handed neutrinos. Note that we do not have any fields which are uncharged under the anomalous $U(1)_{{\rm A}}$ symmetry in the untwisted sector. Thus, this model does not contain any untwisted geometric moduli fields except the dilaton field. The anomaly caused by the anomalous $U(1)_{{\rm A}}$ symmetry can be canceled by the Green-Schwarz mechanism. Mixed anomalies associated with Abelian or non-Abelian groups satisfy the Green-Schwarz universality condition\footnote{ See e.g. \cite{Kobayashi:1996pb} and references therein.},
\begin{align}
\frac{1}{k_a} {\rm Tr}_{G_a} T(R) Q_{{\rm A}}
&=
{\rm Tr} Q_{{\rm B}}^2 Q_{{\rm A}}
=
\frac{1}{3} {\rm Tr} Q_{{\rm A}}^3
=
\frac{1}{24} {\rm Tr} Q_{{\rm A}} 
=
8 \pi^2 \delta_{\rm GS} 
= 
\frac{1}{2} \sqrt{ \frac{5}{3} }.
\end{align}
Here $G_a$ and $k_a$ correspond to a non-Abelian group and its Kac-Moody level. In our case, $k_a =1$ for all non-Abelian groups in this model. $2 T(R)$ represents the Dynkin index of the representation $R$. $Q_{{\rm B}} $ mean charges of the Abelian groups which are orthogonal to the anomalous $U(1)_{{\rm A}}$ symmetry.

Let us proceed to an effective theory analysis to see the structure of the hidden sector. 
The D-term of the anomalous $U(1)_{{\rm A}}$ includes the dilaton-dominant FI-term, 
\begin{equation}
\xi = - \delta_{\rm GS} K_S,
\end{equation}
where $K_S$ is the derivative of the K\"ahler potential by the dilaton $S$.
We can check that the D-flatness conditions for ten $U(1)$s including anomalous $U(1)_{{\rm A}}$ with the FI term, 
\begin{align}
D_{a = 1\ldots 9} &= \sum_i Q_{ai} | f_i |^2 = 0, \qquad
D_A = \sum_i Q_{Ai} | f_i |^2 + \xi= 0, 
\end{align}
can be satisfied if the following fields get non-zero vacuum expectation values (VEVs) as 
\begin{align}
\langle f_4 \rangle &= \langle f_7 \rangle = \langle f_{18} \rangle = \sqrt{2} \left( \frac{3}{5} \right)^{1/4} \sqrt{ \xi}, \nonumber \\
\langle f_{15} \rangle &= \langle f_{17} \rangle = \langle f_{20} \rangle = \langle f_{21} \rangle = \langle f_{22} \rangle = \langle f_{23} \rangle = 
\left( \frac{3}{5} \right)^{1/4} \sqrt{ \xi}.  
\label{VEV}
\end{align}
Note that we omitted the degeneracy factor $D=3$ in the expression, however, their contributions are correctly taken into account. Non-Abelian D-flatness conditions are also satisfied. By the VEVs \eqref{VEV}, $SU(2)^2 \times U(1)^9$ symmetry including the anomalous $U(1)_{{\rm A}}$ symmetry in the original gauge symmetry \eqref{Gaugesymmetry} breaks down, so the unbroken gauge symmetry reads 
\begin{align}
SU(3)_{{\rm C}} \times SU(2)_{{\rm L}}  \times U(1)_Y \times SU(3)^2 \times SU(4).
\label{unbroken}
\end{align}
The tree level superpotential is given by
\begin{align}
W &=
f_1  f_4 f_5 + f_2 f_{10}  f_{11} +  f_3 f_6 f_7 + f_{13} f_{25} f_{27} + f_{14}  f_{44} f_{46} 
+ f_{15} f_{33} f_{36} + f_{16} f_{28} f_{31}  \nonumber \\
&+ f_{17} f_{43} f_{47} + f_{20} f_{34} f_{37} 
+ f_{21} f_{39} f_{41} + f_{22} f_{29} f_{32} + f_{23} f_{40} f_{42} + f_{38} f_{45} f_{48}.
\label{SP}
\end{align}
Here, the degeneracy factor should be taken into account as follows: a three point coupling exists only if all components are different. For instance, for the first term, $f_{1i} f_{4j} f_{5k}$ is allowed only if $i \neq j \neq k$. The coefficients for the three point couplings are of $\mathcal{O} (1)$ and 
they were omitted. This is because the first three terms are untwisted UUU couplings, and the other terms are twisted TTT coupling without world-sheet instanton effect since we make a model at a symmetry enhancement point. Below the string scale, several vectorlike particles become supermassive by the VEVs \eqref{VEV} and decouple. Now we summarize resulting field contents in Table \ref{Tab:MatterContents}. In the visible sector, we have three-generation quarks and leptons of SUSY standard model. Additional fields are vectorlike fields which include up-type and down-type higgs pair, and singlets. We find that we have three hidden sector gauge groups, $SU(3)_1 \times SU(3)_2 \times SU(4)$, with vectorlike flavor numbers $(9,6,6)$ at the tree level superpotential.

\section{Dilaton stabilization}

In this section, we analyze the hidden sector of the effective theory of our model. We analyze the $SU(3)_2 \times SU(4)$ hidden sector  to stabilize the dilaton field, and regard the $SU(3)_1$ hidden sector as SUSY breaking sector.

\subsection{Racetrack potential}

When the anomalous $U(1)_{{\rm A}}$ symmetry breaks down, certain fields develop VEVs and make the anomalous $U(1)_{\rm A}$ vector multiplet $V_{\rm A}$ massive.
Thus, we use the unitary gauge, i.e.,
\begin{equation}
\tilde V_{\rm A} = V_{\rm A} +\frac{1}{q_X} (\ln X +\ln \overline{X} ),
\end{equation}
where $X$ is the chiral field direction eaten by $V_{\rm A}$ and 
$q_X$ is its (effective) charge.
The dilaton $S$ is also non-linearly transformed under $U(1)_{\rm A}$.
Hence, we use the modified dilaton basis,
\begin{equation}
\tilde S = S +\frac{\delta_{{\rm GS}}}{q_X}\ln X.
\end{equation}
For simplicity, hereafter we denote $\tilde S$ as $S$.

We concentrate on the hidden gauge sectors, $SU(3)_2 \times SU(4)$.
The $SU(3)_2$ hidden sector has six flavors, $(Q_{2i}, \overline{Q}_{2i})$.
Also, the $SU(4)$ sector has six flavors, $(Q'_i, \overline{Q}'_i)$ and $(Q''_i, \overline{Q}''_i)$. 
At the level of the superpotential term \eqref{SP}, 
these six flavors are massless.
However, the VEVs (\ref{VEV}) as well as other singlets generate 
their masses through higher dimensional operators,
\begin{align}
\sum_{i,j=1 \ldots 6} m_{2ij} (S'_k ) Q_{2i} \overline{Q}_{2j} 
+ \sum_{i,j=1 \ldots 3} m'_{ij} (S'_k ) Q'_i \overline{Q}'_j 
+ \sum_{i,j=1 \ldots 3} m''_{ij} (S'_k ) Q''_i \overline{Q}''_j .
\end{align}
Here the effective mass matrices $m_{2ij}, m'_{ij}$ and $m''_{ij}$ depend on VEVs of some scalar fields $S'_i$ which are not charged under the unbroken group \eqref{unbroken}.
We assume that all flavors become massive and eigenvalues of the mass matrices are positive, 
however, their mass scales highly depend on the detail of the model.
Hence, we introduce the parameters,
\begin{equation}
m_2 = \frac16 {\rm tr}~(m_{2ij}), \qquad m' = \frac16 \left( {\rm tr}~(m'_{ij})+{\rm tr}~(m''_{ij}) \right).
\end{equation}
If these mass scales are sufficiently high, all the flavors decouple and 
the hidden sectors become pure supersymmetric Yang-Mills theories below such scales.
Then, their gauginos would condensate and induce non-perturbative effects.
That is, the hidden gauge sector $SU(N)$ SQCD including $F$ flavor fields with mass $m$ 
generates the non-perturbative superpotential
\begin{align}
W = 
M_{{\rm s}}^{3-F/N} m^{F/N} e^{ - 8\pi^2 S/N },
\label{SPone}
\end{align}
where the mass scale should be higher than the dynamical scale,  $m > \Lambda_{{\rm c}}=W^{1/3}$.

\begin{figure}[]
  \begin{center}
    \includegraphics[clip,width=8.0cm]{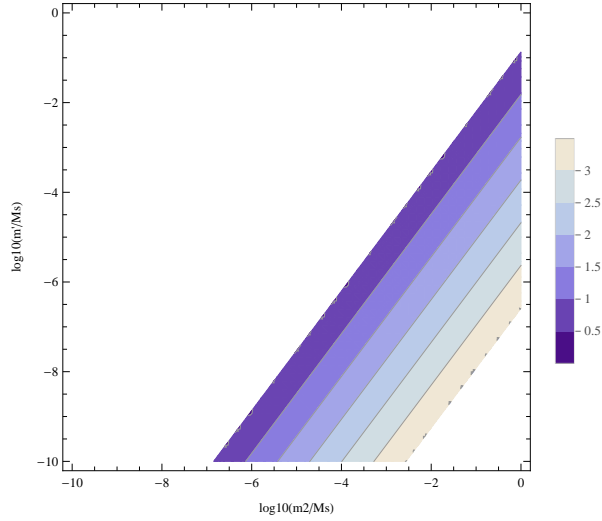}
    \caption{Allowed region for flavor masses $(m_2, m')$. The colored region shows range of the real part of the dilaton VEV, $0.5 \leq {\rm Re}(S) \leq 3.5$.}
    \label{fig:RT}
  \end{center}
\end{figure}

We apply this to the $SU(3)_2 \times SU(4)$ hidden sector.
Then, the following non-perturbative superpotential would be generated, 
\begin{align}
W =
M_{{\rm s}} m_2^2 e^{ -8 \pi^2 S/N_1} 
+ M_{{\rm s}}^{3/2} m'^{3/2} e^{ -8 \pi^2 S/N_2},
\end{align}
where $N_1=3$ and $N_2=4$.
That is the so-called racetrack superpotential.
The stationary point in SUSY preserving vacuum can be evaluated by $D_S W = K_S W + W_S = 0$. 
The real part of the dilaton VEV is approximately given by
\begin{align}
{\rm Re} (S) 
&\approx 
\frac{ 3 \ln (10) }{2 \pi^2}
\left(
\log_{10} \left( \frac{4}{3} \right) 
+ 2 \log_{10} \left( \frac{m_2}{M_{{\rm s}}} \right)
- \frac{3}{2} \log_{10} \left( \frac{m'}{M_{{\rm s}}} \right)
\right).
\end{align}
In Figure \ref{fig:RT}, we show the allowed region for $(m_2, m')$ for the dilaton VEV around  $0.5 \leq {\rm Re}(S) \leq 3.5$. 
For example, the minimal supersymmetric standard model with ${\cal O}(1)$ TeV SUSY breaking scale leads to 
${\rm Re} (S) = 1/g^2 \approx 2$ at the unified scale.
It turns out that we need a hierarchy between $m_2$ and $m'$ of the order of $10^4$ 
to realize a realistic value.

At the above minimum, we can estimate 
\begin{equation}
W \approx M_{{\rm s}}^{3/2} m'^{3/2} e^{ -8 \pi^2 S/N_2}, \qquad 
W_{SS} = \frac{(8 \pi^2)^2 }{N_1 N_2} W.
\end{equation}
Thus, the dilaton mass is much heavier than the gravitino mass by a factor of ${\cal O}(100)$.
Note that, at the stationary point, the vacuum energy has a negative value
\begin{align}
V_0 
&=
- 3 e^{K} | W |^2,
\label{SUSYAdS}
\end{align}
where $W \approx M_{{\rm s}}^{3/2} m'^{3/2} e^{ -8 \pi^2 S/N_2}$.
This vacuum corresponds to SUSY anti de Sitter vacuum.
In the next subsection, we discuss uplifting by the SUSY breaking 
to realize $V_ 0 \approx 0$.

\subsection{SUSY breaking and uplifting}

Here, we discuss the possibility for SUSY breaking and uplifting 
by the remaining hidden $SU(3)_1$ gauge sector.
We have nine flavors of $SU(3)_1$, $(Q_{1i}, \overline{Q}_{1i})$, at the three-point coupling level in the superpotential \eqref{SP}. These fields are also expected to have an effective mass term from higher-dimensional operators, $m_{1ij}(S'_k) Q_{1i} \overline{Q}_{1j} $, and masses are given by a function of VEVs of the scalar fields $S'_i$. 
Their mass spectrum is again highly dependent on the details of the model. 
We assume that four flavors among the nine become light compared to the 
other five flavors, which decouple at low energy.
We denote the light mass as $m_1$.

Under the above assumption, 
the $SU(3)_1$ hidden sector leads to the Intriligator-Seiberg-Shih (ISS)-type of 
metastable SUSY breaking \cite{Intriligator:2006dd}. According to Seiberg duality \cite{Seiberg:1994bz}, $SU(3)$ SQCD with $F=4$ flavors is dual to the supersymmetric system of corresponding mesons $M^i_{j} = Q^i \overline{Q}_j$ and baryons, $B_i = \varepsilon_{ijk\ell}Q^j Q^k Q^\ell$ 
and $ \overline{B}^i = \varepsilon^{ijk\ell}\overline{Q}_j\overline{Q}_k\overline{Q}_\ell$. 
Here $i$ and $j$ run the flavor space, $i, j  = 1 \ldots 4$.  
In the magnetic  theory, SUSY is spontaneously broken. The superpotential is given by
\begin{align}
W &=
\frac{1}{\Lambda_{{\rm d}}^{5}} ( \overline{B}^{T} M B - \det M ) + {\rm Tr} \  m_1 M.
\end{align}
Here $\Lambda_{{\rm d}}$ is a dynamical scale. 
The K\"ahler potential of $M^i_{j}$, $B_i $ and $ \overline{B}^i$ can be written by 
\begin{equation}
K={C_M}^j_{i}\frac{1}{\Lambda_{\rm d}^2}|M^i_{j}|^2 + {C_B}^i\frac{1}{\Lambda_{\rm d}^4}|B_i|^2 +{C_{\overline B}}_i\frac{1}{\Lambda_{\rm d}^4}|\overline B^i|^2,
\end{equation}
where the coefficients, ${C_M}^j_{i}$, ${C_B}^i$ and ${C_{\overline B}}_i$, are of ${\cal O}(1)$.
In the above superpotential, the determinant term is negligible \cite{Intriligator:2006dd}. The F-term associated with $M^i_j$ is given by 
\begin{align}
\frac{\partial W}{\partial M^i_j}  
&=
\frac{1}{\Lambda_{{\rm d}}^{5}} \overline{B}^i B_j + m_1 \delta^i_j, 
\end{align}
and the F-flatness condition for all components of $M^i_j$, $F_{M^i_j} = 0$, cannot be satisfied simultaneously. 
Then SUSY is spontaneously broken. Furthermore, one-loop corrections ensure the stability of the vacuum. 
This SUSY breaking effect has the contribution on the vacuum energy
\begin{align}
V_0'
& \approx \frac{ m_1^2 }{S + \overline S}\Lambda_{{\rm d}}^2,
\label{uplifting}
\end{align}
up to ${\cal O}(1)$ factor. 
The life-time of the metastable vacuum is sufficiently long compared with that of our universe if $m_1 \ll \Lambda_{{\rm d}}$ is satisfied.
Then, we fine-tune the parameters such that the total vacuum energy is almost vanishing, 
$V_0 + V_{0}' \approx 0$.
This  fine-tuning relation leads to 
\begin{align}
\left( \frac{m_1}{M_{{\rm s}}} \right)^2
&=
 {e^{-4 \pi^2 {\rm Re}(S) }} \left( \frac{M_{{\rm s}}}{\Lambda_{{\rm d}}} \right)^2 
\left( \frac{m'}{M_{{\rm s}}} \right)^3,
\end{align}
up to ${\cal O}(1)$ factor.
Note that the dilaton has heavy mass and it is stabilized even with 
the above uplifting potential (see e.g. \cite{Dudas:2006gr}).

In Table \ref{Tab:Parameters}, we show several examples of possible choices of $( m_1, m_2, m' )$ 
satisfying the above relation. We take the string scale $M_{{\rm s}} = 10^{17}\ {\rm GeV}$. SUSY breaking scales are also represented by the gravitino mass $ m_{3/2} \approx F/M_{\rm p}$. In the case of ${\rm Re}(S) \approx 2$, it turns out that the SUSY breaking scale becomes very light since the dynamical scale $\Lambda_{\rm d}$ is very low. However, if we allow the dilaton VEV to be less than or comparable to ${\rm Re}(S) \approx 1$, we can achieve a suitable SUSY breaking scale, $m_{3/2} \approx 1 \ {\rm TeV}$ for instance.   
The range of the dilaton VEV $0.5 < {\rm Re} (S) < 1$ is acceptable since we have several vectorlike pairs in our effective theory as in Table \ref{Tab:MatterContents}. They may also gain masses 
at some energy scale, and 
contribute to the running of gauge coupling constants as threshold corrections.

\section{Conclusion}

We have studied dilaton stabilization by using an explicit model.
In the asymmetric orbifold construction, since we start from a Narain lattice, the resulting model has an enhanced gauge symmetry with rank 22. Thus this model has a relatively large hidden sector. In our explicit model, we have a semi-realistic visible sector and a
$SU(3)_1 \times SU(3)_2 \times SU(4)$ hidden sector. By the ${\bf Z}_3$ asymmetric orbifold action, geometric moduli fields except the dilaton field are projected out in the effective theory. We found an explicit three-generation model with suitable hidden sector for dilaton stabilization by the racetrack mechanism. We also discussed the possibility for ISS-type SUSY breaking and F-term uplifting toward the de Sitter universe. 

The absence of geometric moduli fields and the existence of multiple hidden sectors with relatively large rank are common in other types of asymmetric orbifold constructions, e.g. ${\bf Z}_6$ and ${\bf Z}_{12}$. Regarding the Yukawa structure of our model, this model cannot realize suitable particle masses since the charm quark is as heavy as the top quark mass. This is because our orbifold action is ${\bf Z}_3$, resulting in a quark Yukawa matrix that is far too simple. To address a realistic Yukawa structure, we will consider model building by other types of orbifold actions, e.g. ${\bf Z}_6$, in future work.

Few studies have been carried out on moduli stabilization in explicit and 
realistic models.
Thus, extending our study is important, 
although we have assumed mass scales and used free parameters.
As one of the next issues, it is important to study 
which mass scales can be realized.
For that purpose, we need to study generic higher dimensional operators.
Such studies were done in symmetric orbifold models 
(see e.g. \cite{Kobayashi:2004ud}  and references therein).
We will study its extension to asymmetric orbifold models elsewhere.

%---------------------------------------------------------Acknowledgement

\subsection*{Acknowledgement}
T.K. was supported in part by the Grant-in-Aid for Scientific Research No.~25400252 from the Ministry of Education, Culture, Sports, Science and Technology of Japan.

\begin{table}[t]
\begin{center}
\scriptsize
\begin{tabular}{|c|c|c|cccccccccc|c|c|c|}
\hline
$U/T$ & $f$ & ${\rm Irrep.}$ & $Q_1$ & $Q_2$ & $Q_3$ & $Q_4$ & $Q_5$ & $Q_6$ & $Q_7$ & $Q_8$ & $Q_9$ & $Q_{10}$ 
& $Q_{Y}$ & $Q_{A}$ & ${\rm Deg.}$ \\ 
\hline
 $U$ & $1$ 
& $( {\bf 1},{\bf 1},{\bf 1},{\bf 1},{\bf 1} ,{\bf 1},{\bf 1} )$ & $0$ & $0$ & $0$ & $0$ & $0$ & $0$ & $0$ & $0$ & $1$ & $0$ & $-1$ & $\frac{1}
{4}$ & $3$ \\
 $U$ & $2$  
& $(  {\bf 1},{\bf 1},{\bf 1},{\bf 1},{\bf 1},{\bf 1},{\bf 1} )$ 
& $0$ & $0$ & $0$ & $0$ & $0$ & $0$ & $1$ & $0$ & $0$ & $0$ & $0$ & $-\frac{1}{2}$ & $3$ \\
 $U$ & $3$  
& $( {\bf 1},{\bf 1},{\bf 1},{\bf 1},{\bf 1},{\bf 1},{\bf 1} )$ 
& $0$ & $0$ & $0$ & $0$ & $1$ & $0$ & $0$ & $0$ & $0$ & $0$ & $-1$ & $\frac{1}{4}$ & $3$ \\
$ U$ & $4$  
& $( {\bf 1},{\bf 1},{\bf 1},{\bf 1},{\bf 1},{\bf 1},{\bf 2} )$ 
& $0$ & $0$ & $0$ & $0$ & $0$ & $0$ & $0$ & $-1$ & $-\frac{1}{2}$ & $0$ & $0$ & $-\frac{1}{2}$ & $3$ \\
 $U$ & $5$  
& $({\bf 1},{\bf 1},{\bf 1},{\bf 1},{\bf 1},{\bf 1},{\bf 2} )$ 
& $0$ & $0$ & $0$ & $0$ & $0$ & $0$ & $0$ & $1$ & $-\frac{1}{2}$ & $0$ & $1$ & $\frac{1}{4}$ & $3$ \\
 $U$ & $6$  
& $({\bf 1},{\bf 1},{\bf 1},{\bf 1},{\bf 1},{\bf 2},{\bf 1} )$ 
& $0$ & $0$ & $0$ & $-1$ & $-\frac{1}{2}$ & $0$ & $0$ & $0$ & $0$ & $0$ & $1$ & $\frac{1}{4}$ & $3$ \\
 $U$ & $7$  
& $({\bf 1},{\bf 1},{\bf 1},{\bf 1},{\bf 1},{\bf 2},{\bf 1} )$ 
& $0$ & $0$ & $0$ & $1$ & $-\frac{1}{2}$ & $0$ & $0$ & $0$ & $0$ & $0$ & $0$ & $-\frac{1}{2}$ & $3$ \\
 $U$ & $8$ 
& $( {\bf 1},{\bf 1},{\bf 1},\overline{{\bf 3}},{\bf 1},{\bf 1},{\bf 1} )$ 
& $0$ & $-1$ & $0$ & $0$ & $0$ & $0$ & $0$ & $0$ & $0$ & $0$ & $0$ & $\frac{1}{4}$ & $3$ \\
 $U$ & $9$ 
& $({\bf 1},{\bf 1},\overline{{\bf 3}},{\bf 1},{\bf 1},{\bf 1},{\bf 1} ) $
& $0$ & $0$ & $-1$ & $0$ & $0$ & $0$ & $0$ & $0$ & $0$ & $0$ & $0$ & $\frac{1}{4}$ & $3$ \\
 $U$ & $10$  
& $( {\bf 1},{\bf 2},{\bf 1},{\bf 1},{\bf 1},{\bf 1},{\bf 1} )$ 
& $0$ & $0$ & $0$ & $0$ & $0$ & $-1$ & $-\frac{1}{2}$ & $0$ & $0$ & $0$ & $-\frac{1}{2}$ & $1$ & $3$ \\
 $U$ & $11$  
& $( {\bf 1},{\bf 2},{\bf 1},{\bf 1},{\bf 1},{\bf 1},{\bf 1} ) $
& $0$ & $0$ & $0$ & $0$ & $0$ & $1$ & $-\frac{1}{2}$ & $0$ &$ 0$ & $0$ & $\frac{1}{2}$ & $-\frac{1}{2}$ & $3$ \\
$U$ & $12$  
& $( \overline{{\bf 3}},{\bf 1},{\bf 1},{\bf 1},{\bf 1},{\bf 1},{\bf 1} ) $
& $1$ & $0$ & $0$ & $0$ & $0$ & $0$ & $0$ & $0$ & $0$ & $0$ & $\frac{1}{3}$ & $1$ & $3$ \\
 $T$ & $13$  
& $( {\bf 1},{\bf 1},{\bf 1},{\bf 1},{\bf 1},{\bf 1},{\bf 1} ) $
& $-\frac{1}{2}$ & $-\frac{1}{4}$ & $-\frac{1}{4}$ & $\frac{1}{2}$ & $\frac{1}{6}$ & $0$ & $\frac{2}{3}$ & $-\frac{1}{2}$ & $\frac{1}{6}$ & $0$ & 
$-1$ & $-1$ & $3$ \\
 $T$ & $14$  
& $({\bf 1},{\bf 1},{\bf 1},{\bf 1},{\bf 1},{\bf 1},{\bf 1})$ 
& $\frac{1}{4}$ & $-\frac{1}{4}$ & $-\frac{1}{4}$ & $0$ & $-\frac{1}{3}$ & $\frac{1}{2}$ & $\frac{1}{6}$ & $0$ & $\frac{2}{3}$ & $-1$ & $0$ & $0$ 
& $3$ \\
 $T$ & $15$  
& $({\bf 1},{\bf 1},{\bf 1},{\bf 1},{\bf 1},{\bf 1},{\bf 1} )$ 
& $\frac{1}{4}$ & $-\frac{1}{4}$ & $-\frac{1}{4}$ & $0$ & $-\frac{1}{3}$ & $\frac{1}{2}$ & $\frac{1}{6}$ & $0$ & $\frac{2}{3}$ & $1$ & $0$ & $0$ 
& $3$ \\
 $T$ & $16$ 
& $( {\bf 1},{\bf 1},{\bf 1},{\bf 1},{\bf 1},{\bf 1},{\bf 1} )$
& $\frac{1}{4}$ & $-\frac{1}{4}$ & $-\frac{1}{4}$ & $0$ & $\frac{2}{3}$ & $\frac{1}{2}$ & $\frac{1}{6}$ & $0$ & $-\frac{1}{3}$ & $-1$ & $0$ & $0$ 
& $3$ \\
 $T$ & $17$ 
& $({\bf 1},{\bf 1},{\bf 1},{\bf 1},{\bf 1},{\bf 1},{\bf 1} ) $
& $\frac{1}{4}$ & $-\frac{1}{4}$ & $-\frac{1}{4}$ & $0$ & $\frac{2}{3}$ & $\frac{1}{2}$ & $\frac{1}{6}$ & $0$ & $-\frac{1}{3}$ & $1$ & $0$ & $0$ 
& $3$ \\
 $T$ & $18$ 
& $({\bf 1},{\bf 1},{\bf 1},{\bf 1},{\bf 1},{\bf 1},{\bf 1} ) $
& $\frac{1}{4}$ & $\frac{1}{2}$ & $\frac{1}{2}$ & $-\frac{1}{2}$ & $\frac{1}{6}$ & $-\frac{1}{2}$ & $\frac{1}{6}$ & $\frac{1}{2}$ & $\frac{1}{6}$ 
& $-1$ & $0$ & $\frac{3}{4}$ & $3$ \\
 $T$ & $19$ 
& $({\bf 1},{\bf 1},{\bf 1},{\bf 1},{\bf 1},{\bf 1},{\bf 1} ) $
& $\frac{1}{4}$ & $\frac{1}{2}$ & $\frac{1}{2}$ & $-\frac{1}{2}$ & $\frac{1}{6}$ & $-\frac{1}{2}$ & $\frac{1}{6}$ & $\frac{1}{2}$ & $\frac{1}{6}$ 
& $1$ & $0$ & $\frac{3}{4}$ & $3$ \\
 $T$ & $20$ 
& $({\bf 1},{\bf 1},{\bf 1},{\bf 1},{\bf 1},{\bf 1},{\bf 2} ) $
& $-\frac{1}{2}$ & $-\frac{1}{4}$ & $-\frac{1}{4}$ & $\frac{1}{2}$ & $\frac{1}{6}$ & $0$ & $-\frac{1}{3}$ & $\frac{1}{2}$ & $-\frac{1}{3}$ & $0$ 
& $0$ & $-\frac{1}{4}$ & $3$ \\
 $T$ & $21$ 
& $( {\bf 1},{\bf 1},{\bf 1},{\bf 1},{\bf 1},{\bf 1},{\bf 2} ) $
& $-\frac{1}{2}$ & $-\frac{1}{4}$ & $\frac{1}{2}$ & $-\frac{1}{2}$ & $\frac{1}{6}$ & $\frac{1}{2}$ & $\frac{1}{6}$ & $0$ & $\frac{1}{6}$ & $
\frac{1}{2}$ & $0$ & $-\frac{3}{4}$ & $3$ \\
 $T$ & $22$ 
& $( {\bf 1},{\bf 1},{\bf 1},{\bf 1},{\bf 1},{\bf 2},{\bf 1} )$ 
& $-\frac{1}{2}$ & $-\frac{1}{4}$ & $-\frac{1}{4}$ & $-\frac{1}{2}$ & $-\frac{1}{3}$ & $0$ & $-\frac{1}{3}$ & $-\frac{1}{2}$ & $\frac{1}{6}$ & 
$0$ & $0$ & $-\frac{1}{4}$ & $3$ \\
 $T$ & $23$ 
& $( {\bf 1},{\bf 1},{\bf 1},{\bf 1},{\bf 1},{\bf 2},{\bf 1} ) $
& $-\frac{1}{2}$ & $\frac{1}{2}$ & $-\frac{1}{4}$ & $0$ & $\frac{1}{6}$ & $\frac{1}{2}$ & $\frac{1}{6}$ & $\frac{1}{2}$ & $\frac{1}{6}$ & $-
\frac{1}{2}$ & $0$ & $-\frac{3}{4}$ & $3$ \\
 $T$ & $24$ 
& $({\bf 1},{\bf 1},{\bf 1},{\bf 1},{\bf 4},{\bf 1},{\bf 1} ) $
& $\frac{1}{4}$ & $-\frac{1}{4}$ & $-\frac{1}{4}$ & $\frac{1}{2}$ & $\frac{1}{6}$ & $-\frac{1}{2}$ & $\frac{1}{6}$ & $\frac{1}{2}$ & $\frac{1}
{6}$ & $-\frac{1}{2}$ & $-\frac{1}{2}$ & $\frac{3}{4}$ & $3$ \\
 $T$ & $25$ 
& $({\bf 1},{\bf 1},{\bf 1},{\bf 1},{\bf 4},{\bf 1},{\bf 1} ) $
& $\frac{1}{4}$ & $-\frac{1}{4}$ & $\frac{1}{2}$ & $-\frac{1}{2}$ & $\frac{1}{6}$ & $0$ & $-\frac{1}{3}$ & $0$ & $-\frac{1}{3}$ & $0$ & $\frac
{1}{2}$ & $\frac{1}{2}$ & $3$ \\
 $T$ & $26$ 
& $( {\bf 1},{\bf 1},{\bf 1},{\bf 1},\overline{\bf 4},{\bf 1},{\bf 1} ) $
& $\frac{1}{4}$ & $-\frac{1}{4}$ & $-\frac{1}{4}$ & $-\frac{1}{2}$ & $\frac{1}{6}$ & $-\frac{1}{2}$ & $\frac{1}{6}$ & $-\frac{1}{2}$ & $\frac{1}
{6}$ & $\frac{1}{2}$ & $-\frac{1}{2}$ & $\frac{3}{4}$ & $3$ \\
 $T$ & $27$ 
& $( {\bf 1},{\bf 1},{\bf 1},{\bf 1},\overline{\bf 4},{\bf 1},{\bf 1} ) $
& $\frac{1}{4}$ & $\frac{1}{2}$ & $-\frac{1}{4}$ & $0$ & $-\frac{1}{3}$ & $0$ & $-\frac{1}{3}$ & $\frac{1}{2}$ & $\frac{1}{6}$ & $0$ & $\frac{1}
{2}$ & $\frac{1}{2}$ & $3$ \\
 $T$ & $28$ 
& $({\bf 1},{\bf 1},{\bf 1},{\bf 3},{\bf 1},{\bf 1},{\bf 1} ) $ 
& $-\frac{1}{2}$ & $0$ & $-\frac{1}{4}$ & $0$ & $-\frac{1}{3}$ & $-\frac{1}{2}$ & $\frac{1}{6}$ & $\frac{1}{2}$ & $\frac{1}{6}$ & $\frac{1}{2}$ & 
$0$ & $0$ & $3$ \\
 $T$ & $29$ 
& $({\bf 1},{\bf 1},{\bf 1},{\bf 3},{\bf 1},{\bf 1},{\bf 1} ) $
& $\frac{1}{4}$ & $0$ & $\frac{1}{2}$ & $\frac{1}{2}$ & $\frac{1}{6}$ & $\frac{1}{2}$ & $\frac{1}{6}$ & $\frac{1}{2}$ & $\frac{1}{6}$ & $0$ & $0$ 
& $-\frac{1}{4}$ & $3$ \\
 $T$ & $30$ 
& $( {\bf 1},{\bf 1},{\bf 1},{\bf 3},{\bf 1},{\bf 1},{\bf 2} ) $
& $\frac{1}{4}$ & $0$ & $-\frac{1}{4}$ & $-\frac{1}{2}$ & $\frac{1}{6}$ & $0$ & $-\frac{1}{3}$ & $0$ & $\frac{1}{6}$ & $-\frac{1}{2}$ & $0$ & $
\frac{3}{4}$ & $3$ \\
 $T$ & $31$ 
& $( {\bf 1},{\bf 1},{\bf 1},\overline{\bf 3},{\bf 1},{\bf 1},{\bf 1} ) $
& $\frac{1}{4}$ & $\frac{1}{4}$ & $\frac{1}{2}$ & $0$ & $-\frac{1}{3}$ & $0$ & $-\frac{1}{3}$ & $-\frac{1}{2}$ & $\frac{1}{6}$ & $\frac{1}{2}$ & 
$0$ & $0$ & $3$ \\
 $T$ & $32$ 
& $({\bf 1},{\bf 1},{\bf 1},\overline{\bf 3},{\bf 1},{\bf 2},{\bf 1} )$ 
& $\frac{1}{4}$ & $\frac{1}{4}$ & $-\frac{1}{4}$ & $0$ & $\frac{1}{6}$ & $-\frac{1}{2}$ & $\frac{1}{6}$ & $0$ & $-\frac{1}{3}$ & $0$ & $0$ & $
\frac{1}{2}$ & $3$ \\
 $T$ & $33$ 
& $({\bf 1},{\bf 1},{\bf 3},{\bf 1},{\bf 1},{\bf 1},{\bf 1} )$
& $-\frac{1}{2}$ & $-\frac{1}{4}$ & $0$ & $-\frac{1}{2}$ & $\frac{1}{6}$ & $-\frac{1}{2}$ & $\frac{1}{6}$ & $0$ & $-\frac{1}{3}$ & $-\frac{1}{2}$ 
& $0$ & $0$ & $3$ \\
 $T$ & $34$ 
& $({\bf 1},{\bf 1},{\bf 3},{\bf 1},{\bf 1},{\bf 1},{\bf 1} ) $
& $\frac{1}{4}$ & $\frac{1}{2}$ & $0$ & $-\frac{1}{2}$ & $\frac{1}{6}$ & $\frac{1}{2}$ & $\frac{1}{6}$ & $-\frac{1}{2}$ & $\frac{1}{6}$ & $0$ & 
$0$ & $-\frac{1}{4}$ & $3$ \\
 $T$ & $35$ 
& $({\bf 1},{\bf 1},{\bf 3},{\bf 1},{\bf 1},{\bf 2},{\bf 1} ) $
& $\frac{1}{4}$ & $-\frac{1}{4}$ & $0$ & $0$ & $\frac{1}{6}$ & $0$ & $-\frac{1}{3}$ & $\frac{1}{2}$ & $\frac{1}{6}$ & $\frac{1}{2}$ & $0$ & $
\frac{3}{4}$ & $3$ \\
 $T$ & $36$ 
& $({\bf 1},{\bf 1},\overline{\bf 3},{\bf 1},{\bf 1},{\bf 1},{\bf 1} )$ 
& $\frac{1}{4}$ & $\frac{1}{2}$ & $\frac{1}{4}$ & $\frac{1}{2}$ & $\frac{1}{6}$ & $0$ & $-\frac{1}{3}$ & $0$ & $-\frac{1}{3}$ & $-\frac{1}{2}$ & 
$0$ & $0$ & $3$ \\
 $T$ & $37$ 
& $({\bf 1},{\bf 1},\overline{\bf 3},{\bf 1},{\bf 1},{\bf 1},{\bf 2} )$
& $\frac{1}{4}$ & $-\frac{1}{4}$ & $\frac{1}{4}$ & $0$ & $-\frac{1}{3}$ & $-\frac{1}{2}$ & $\frac{1}{6}$ & $0$ & $\frac{1}{6}$ & $0$ & $0$ & $
\frac{1}{2}$ & $3$ \\
 $T$ & $38$ 
& $( {\bf 1},{\bf 2},{\bf 1},{\bf 1},{\bf 1},{\bf 1},{\bf 1} ) $
& $-\frac{1}{2}$ & $\frac{1}{2}$ & $\frac{1}{2}$ & $0$ & $-\frac{1}{3}$ & $0$ & $\frac{1}{6}$ & $0$ & $-\frac{1}{3}$ & $0$ & $\frac{1}{2}$ & $-1$ 
& $3$ \\
 $T$ & $39$ 
& $( {\bf 1},{\bf 2},{\bf 1},{\bf 1},{\bf 1},{\bf 1},{\bf 1} )$ 
& $\frac{1}{4}$ & $-\frac{1}{4}$ & $-\frac{1}{4}$ & $0$ & $-\frac{1}{3}$ & $-\frac{1}{2}$ & $-\frac{1}{3}$ & $0$ & $-\frac{1}{3}$ & $-1$ & $
\frac{1}{2}$ & $\frac{3}{4}$ & $3$ \\
 $T$ & $40$ 
& $( {\bf 1},{\bf 2},{\bf 1},{\bf 1},{\bf 1},{\bf 1},{\bf 1} )$ 
& $\frac{1}{4}$ & $-\frac{1}{4}$ & $-\frac{1}{4}$ & $0$ & $-\frac{1}{3}$ & $-\frac{1}{2}$ & $-\frac{1}{3}$ & $0$ & $-\frac{1}{3}$ & $1$ & $\frac
{1}{2}$ & $\frac{3}{4}$ & $3$ \\
 $T$ & $41$ 
& $({\bf 1},{\bf 2},{\bf 1},{\bf 1},{\bf 1},{\bf 1},{\bf 2} ) $
& $\frac{1}{4}$ & $\frac{1}{2}$ & $-\frac{1}{4}$ & $\frac{1}{2}$ & $\frac{1}{6}$ & $0$ & $\frac{1}{6}$ & $0$ & $\frac{1}{6}$ & $\frac{1}{2}$ & 
$-\frac{1}{2}$ & $0$ & $3$ \\
 $T$ & $42$ 
& $({\bf 1},{\bf 2},{\bf 1},{\bf 1},{\bf 1},{\bf 2},{\bf 1} )$ 
& $\frac{1}{4}$ & $-\frac{1}{4}$ & $\frac{1}{2}$ & $0$ & $\frac{1}{6}$ & $0$ & $\frac{1}{6}$ & $-\frac{1}{2}$ & $\frac{1}{6}$ & $-\frac{1}{2}$ & 
$-\frac{1}{2}$ & $0$ & $3$ \\
 $T$ & $43$ 
& $(\overline{\bf 3},{\bf 1},{\bf 1},{\bf 1},{\bf 1},{\bf 1},{\bf 1} ) $
& $-\frac{1}{4}$ & $-\frac{1}{4}$ & $\frac{1}{2}$ & $0$ & $-\frac{1}{3}$ & $0$ & $-\frac{1}{3}$ & $\frac{1}{2}$ & $\frac{1}{6}$ & $-\frac{1}{2}$ 
& $\frac{1}{3}$ & $0$ & $3$ \\
 $T$ & $44$ 
& $(\overline{\bf 3},{\bf 1},{\bf 1},{\bf 1},{\bf 1},{\bf 1},{\bf 1} ) $
& $-\frac{1}{4}$ & $\frac{1}{2}$ & $-\frac{1}{4}$ & $-\frac{1}{2}$ & $\frac{1}{6}$ & $0$ & $-\frac{1}{3}$ & $0$ & $-\frac{1}{3}$ & $\frac{1}{2}$ 
& $\frac{1}{3}$ & $0$ & $3$ \\
 $T$ & $45$ 
& $(\overline{\bf 3},{\bf 1},{\bf 1},{\bf 1},{\bf 1},{\bf 1},{\bf 1 } ) $
& $\frac{1}{2}$ & $-\frac{1}{4}$ & $-\frac{1}{4}$ & $\frac{1}{2}$ & $\frac{1}{6}$ & $0$ & $-\frac{1}{3}$ & $-\frac{1}{2}$ & $\frac{1}{6}$ & $0$ & 
$-\frac{2}{3}$ & $\frac{1}{2}$ & $3$ \\
 $T$ & $46$ 
& $({\bf 3},{\bf 1},{\bf 1},{\bf 1},{\bf 1},{\bf 1},{\bf 1} ) $
& $0$ & $-\frac{1}{4}$ & $\frac{1}{2}$ & $\frac{1}{2}$ & $\frac{1}{6}$ & $-\frac{1}{2}$ & $\frac{1}{6}$ & $0$ & $-\frac{1}{3}$ & $\frac{1}{2}$ & 
$-\frac{1}{3}$ & $0$ & $3$ \\
 $T$ & $47$ 
& $({\bf 3},{\bf 1},{\bf 1},{\bf 1},{\bf 1},{\bf 1},{\bf 1} ) $
& $0$ & $\frac{1}{2}$ & $-\frac{1}{4}$ & $0$ & $-\frac{1}{3}$ & $-\frac{1}{2}$ & $\frac{1}{6}$ & $-\frac{1}{2}$ & $\frac{1}{6}$ & $-\frac{1}{2}$ 
& $-\frac{1}{3}$ & $0$ & $3$ \\
 $T$ & $48$ 
& $({\bf 3},{\bf 2},{\bf 1},{\bf 1},{\bf 1},{\bf 1},{\bf 1} ) $
& $0$ & $-\frac{1}{4}$ & $-\frac{1}{4}$ & $-\frac{1}{2}$ & $\frac{1}{6}$ & $0$ & $\frac{1}{6}$ & $\frac{1}{2}$ & $\frac{1}{6}$ & $0$ & $\frac{1}
{6}$ & $\frac{1}{2}$ & $3$ \\
\hline
\end{tabular}
\caption[smallcaption]{Massless spectrum of three-generation $SU(3)_{{\rm C}} \times SU(2)_{{\rm L}} \times U(1)_{{\rm Y}}$ model. Representations under the non-Abelian group $SU(3)_{{\rm C}} \times SU(2)_{{\rm L}} \times SU(3)^2 \times SU(4) \times SU(2)^2$ and $U(1)$ charges are listed. U and T denote the untwisted and twisted sector respectively. Every field has the degeneracy 3. The gravity and gauge supermultiplets are omitted. The last $SU(2)^2$ and $U(1)^9$ groups are broken by the D-flatness condition.}
\label{Tab:Spectrum}
\end{center}
\end{table}

\begin{table}[t]
\begin{center}
%\scriptsize
\begin{tabular}{|c|l|c||c|l|c|}
\hline
 {\rm Label} & ${\rm Irrep.}$ & $\#$ &  {\rm Label} & ${\rm Irrep.}$ & $\#$ \\ 
\hline
 $Q_i$ & $( {\bf 3}, {\bf 2}, {\bf 1}, {\bf 1}, {\bf 1})_{1/6}$ & $3$ 
&  $\overline{U}_i$ & $( \overline{{\bf 3}}, {\bf 1}, {\bf 1}, {\bf 1}, {\bf 1})_{-2/3}$ & $3$ \\
 $\overline{D}_i$ & $( \overline{{\bf 3}}, {\bf 1}, {\bf 1}, {\bf 1}, {\bf 1})_{1/3}$ & $6$ 
&  $D_i$ & $( {\bf 3}, {\bf 1}, {\bf 1}, {\bf 1}, {\bf 1})_{-1/3}$ & $3$ \\
 $L_i$ & $( {\bf 1}, {\bf 2}, {\bf 1}, {\bf 1}, {\bf 1})_{-1/2}$ & $9$ 
&  $\overline{L}_i$ & $( {\bf 1}, {\bf 2}, {\bf 1}, {\bf 1}, {\bf 1})_{1/2}$ & $6$ \\
 $\overline{E}_i$ & $( {\bf 1}, {\bf 1}, {\bf 1}, {\bf 1}, {\bf 1})_{1}$ & $6$ 
&  $E_i$ & $( {\bf 1}, {\bf 1}, {\bf 1}, {\bf 1}, {\bf 1})_{-1}$ & $3$ \\
 $Q_{1i}$ & $( {\bf 1}, {\bf 1}, {\bf 3}, {\bf 1}, {\bf 1})_{0}$ & $9$ 
&  $\overline{Q}_{1i}$ & $( {\bf 1}, {\bf 1}, \overline{{\bf 3}}, {\bf 1}, {\bf 1})_{0}$ & $9$ \\
 $Q_{2i}$ & $( {\bf 1}, {\bf 1}, {\bf 1}, {\bf 3}, {\bf 1})_{0}$ & $6$ 
&  $\overline{Q}_{2i}$ & $( {\bf 1}, {\bf 1}, {\bf 1}, \overline{{\bf 3}}, {\bf 1})_{0}$ & $6$ \\
 $Q'_i$ & $( {\bf 1}, {\bf 1}, {\bf 1}, {\bf 1}, {\bf 4})_{-1/2}$ & $3$ 
&  $\overline{Q}'_i$ & $( {\bf 1}, {\bf 1}, {\bf 1}, {\bf 1}, \overline{{\bf 4}})_{1/2}$ & $3$ \\
 $Q''_i$ & $( {\bf 1}, {\bf 1}, {\bf 1}, {\bf 1}, {\bf 4})_{1/2}$ & $3$ 
&  $\overline{Q}''_i$ & $( {\bf 1}, {\bf 1}, {\bf 1}, {\bf 1}, \overline{{\bf 4}})_{-1/2}$ & $3$ \\
 $S_i$ & $( {\bf 1}, {\bf 1}, {\bf 1}, {\bf 1}, {\bf 1})_{0}$ & $12$ 
&   & & \\
\hline
\end{tabular}
\caption[smallcaption]{Field contents of effective theory of three-generation model. Representations under the unbroken group $SU(3)_{{\rm C}} \times SU(2)_{{\rm L}} \times SU(3)_1 \times SU(3)_2 \times SU(4) \times U(1)_{{\rm Y}}$ are listed.}
\label{Tab:MatterContents}
\end{center}
\end{table}

\begin{table}[t]
\begin{center}
%\scriptsize
\begin{tabular}{|c|c|c|c|c|c|c|c|}
\hline
 $\log_{10}m_1$ & $\log_{10}m_2$ &  $\log_{10}m'$ & ${\rm Re}(S)$ & ${\rm Im}(S)$ & $\log_{10}\Lambda_{{\rm c}}$ 
& $\log_{10}\Lambda_{{\rm d}}$ & $\log_{10}m_{3/2}$ \\ 
\hline
 $-8.7$ & $-2.7$ & $-5.0$ & $1.0$ & $-3/2\pi$ & $-6.0$ & $-8.2$ & $-16.9$ \\
 $-9.7$ & $-3.2$ & $-5.7$ & $1.0$ & $-3/2\pi$ & $-6.4$ & $-8.2$ & $-17.9$ \\
 $-7.2$ & $-2.5$ & $-4.3$ & $0.8$ & $-3/2\pi$ & $-5.1$ & $-6.9$ & $-14.1$ \\
 $-9.2$ & $-3.5$ & $-5.6$ & $0.8$ & $-3/2\pi$ & $-5.8$ & $-6.9$ & $-16.1$ \\
 $-5.2$ & $-2.2$ & $-3.4$ & $0.5$ & $-3/2\pi$ & $-3.8$ & $-4.8$ & $-10.1$ \\
 $-6.2$ & $-2.7$ & $-4.0$ & $0.5$ & $-3/2\pi$ & $-4.1$ & $-4.8$ & $-11.1$ \\
\hline
\end{tabular}
\caption[smallcaption]{Several choices of flavor masses $(m_1, m_2, m' )$. Quantities with mass dimension are expressed in the planck unit $\log_{10} (m/M_{\rm p})$. 
}
\label{Tab:Parameters}
\end{center}
\end{table}

\end{document}